\documentclass[twocolumn]{aastex631}

\usepackage{longtable}
\usepackage[flushleft]{threeparttable}
\usepackage{tabularx}
\usepackage{multirow}
\usepackage{graphicx}
\usepackage{amsmath,amssymb}
\usepackage{color}
\usepackage{units}
\usepackage{epstopdf}
\usepackage{hyperref}
\usepackage{multirow}
\usepackage{url}
\usepackage{subfigure}
\usepackage{rotating}
\usepackage{amsmath}
\usepackage{enumitem}\setlist[description]{font=\textendash\enskip\scshape\bfseries}

\usepackage[colorinlistoftodos]{todonotes}

\usepackage{tikz}
\usetikzlibrary{shapes.callouts, shadows, arrows,
  positioning,
  shapes.geometric,
  shapes,
  fit}

\newcommand{\beq}{\begin{equation}}
\newcommand{\eeq}{\end{equation}}
\newcommand{\bdm}{\begin{displaymath}}
\newcommand{\edm}{\end{displaymath}}
\newcommand{\T}[1]{\tilde{#1}}

\definecolor{Gray}{gray}{0.9}
\definecolor{orange}{rgb}{0.9,0.5,0}

\newcommand{\boom}{\texttt{BOOM }}
\newcommand{\babamul}{\texttt{Babamul }}
\newcommand{\kowalski}{\texttt{Kowalski }}
\newcommand{\kafka}{\texttt{Kafka }}
\newcommand{\valkey}{\texttt{Valkey }}
\newcommand{\mongo}{\texttt{MongoDB }}
\newcommand{\skyportal}{\texttt{SkyPortal }}

\newcommand{\boomthroughputfactor}{7}

\newcommand{\boommaxmem}{1}
\newcommand{\kowalskimaxmem}{13}

\newcommand{\refconfignworkers}{8}

\begin{document}

\title{\boom and \texttt{Babamul}: a real-time, multi-survey, optical alert broker system operating at scale}
\journalinfo{}

\author[0009-0003-6181-4526]{Theophile Jegou du Laz}
\affil{Division of Physics, Mathematics, and Astronomy, California Institute of Technology, Pasadena, CA 91125, USA}

\correspondingauthor{Theophile Jegou du Laz}
\email{tdulaz@caltech.edu}

\author[0000-0002-8262-2924]{Michael W. Coughlin}
\affil{School of Physics and Astronomy, University of Minnesota, Minneapolis, Minnesota 55455, USA}

\author[0000-0003-4039-6954]{Peter Bachant}
\affil{Division of Physics, Mathematics, and Astronomy, California Institute of Technology, Pasadena, CA 91125, USA}

\author{Jacob E. Simones}
\affil{School of Physics and Astronomy, University of Minnesota, Minneapolis, Minnesota 55455, USA}
\affil{Department of Physics, University of Minnesota-Duluth Duluth MN 55812 USA}

\author[0009-0008-3603-0013]{Thomas Culino}
\affil{Division of Physics, Mathematics, and Astronomy, California Institute of Technology, Pasadena, CA 91125, USA}

\author[0009-0009-7000-8343]{Antoine Le Calloch}
\affil{School of Physics and Astronomy, University of Minnesota, Minneapolis, Minnesota 55455, USA}

\author[0000-0003-1314-4241]{Sushant Sharma Chaudhary}
\affil{School of Physics and Astronomy, University of Minnesota, Minneapolis, Minnesota 55455, USA}

\author[0000-0002-9364-5419]{Xander J. Hall}
\affil{McWilliams Center for Cosmology and Astrophysics, Department of Physics, Carnegie Mellon University, 5000 Forbes Avenue, Pittsburgh, PA 15213 }

\author[0000-0002-4843-345X]{Tyler Barna}
\affil{School of Physics and Astronomy, University of Minnesota, Minneapolis, Minnesota 55455, USA}

\author{Daniel Warshofsky}
\affiliation{School of Physics and Astronomy, University of Minnesota, Minneapolis, Minnesota 55455, USA}

\author[0000-0002-3168-0139]{Matthew Graham}
\affil{Division of Physics, Mathematics, and Astronomy, California Institute of Technology, Pasadena, CA 91125, USA}

\author[0000-0002-5619-4938]{Mansi M. Kasliwal}
\affil{Division of Physics, Mathematics, and Astronomy, California Institute of Technology, Pasadena, CA 91125, USA}

\author[0000-0003-2242-0244]{Ashish Mahabal}
\affil{Division of Physics, Mathematics, and Astronomy, California Institute of Technology, Pasadena, CA 91125, USA}

\author[0000-0002-7777-216X]{Joshua S. Bloom}
\affil{Department of Astronomy, University of California, Berkeley, CA 94720, USA}
\affil{Lawrence Berkeley National Laboratory, 1 Cyclotron Road, MS 50B-4206, Berkeley, CA 94720, USA}

\author[0000-0002-6011-0530]{Antonella Palmese}
\affil{McWilliams Center for Cosmology and Astrophysics, Department of Physics, Carnegie Mellon University, 5000 Forbes Avenue, Pittsburgh, PA 15213 }

\author[0000-0002-8532-9395]{Frank J. Masci}
\affiliation{IPAC, California Institute of Technology, 1200 E. California
             Blvd, Pasadena, CA 91125, USA}

\author[0000-0001-5668-3507]{Steven L. Groom}
\affiliation{IPAC, California Institute of Technology, 1200 E. California
             Blvd, Pasadena, CA 91125, USA}

\author[0000-0002-5884-7867]{Richard Dekany}
\affiliation{Caltech Optical Observatories, California Institute of Technology, Pasadena, CA 91125}

\author[0000-0002-0387-370X]{Reed L. Riddle}
\affiliation{Division of Physics, Mathematics, and Astronomy, California Institute of Technology, Pasadena, CA 91125, USA}

\author[0000-0003-3367-3415]{George Helou}
\affiliation{IPAC, California Institute of Technology, 1200 E. California
             Blvd, Pasadena, CA 91125, USA}

\begin{abstract}
With the arrival of ever higher throughput wide-field surveys and a multitude of multi-messenger and multi-wavelength instruments to complement them, software capable of harnessing these associated data streams is urgently required. To meet these needs, a number of community supported \emph{alert brokers} have been built, currently focused on processing of Zwicky Transient Facility (ZTF; $\sim 10^5$--$10^6$ alerts per night) with an eye towards Vera C. Rubin Observatory's Legacy Survey of Space and Time (LSST; $\sim 2 \times 10^7$ alerts per night). Building upon the system that successfully ran in production for ZTF's first seven years of operation, we introduce \texttt{BOOM} (Burst \& Outburst Observations Monitor), an analysis framework focused on real-time, joint brokering of these alert streams.
\texttt{BOOM} harnesses the performance of a \texttt{Rust}-based software stack relying on a non-relational \texttt{MongoDB} database combined with a \texttt{Valkey} in-memory processing queue and a \texttt{Kafka} cluster for message sharing.
With this system, we demonstrate feature parity with the existing ZTF system with a throughput $\sim \boomthroughputfactor \times$ higher.
We describe the workflow that enables the real-time processing as well as the results with custom filters we have built to demonstrate the system's capabilities.
In conclusion, we present the development roadmap for both \texttt{BOOM} and \texttt{Babamul}---the public-facing LSST alert broker built atop \texttt{BOOM}---as we begin the Rubin era.
\end{abstract}

%%%%%%%%%%%%%%%%% BODY OF PAPER %%%%%%%%%%%%%%%%%%

\section{Introduction}

Today, modern optical surveys scan the entire sky daily, reaching depths that allow detection of both distant, bright objects and nearby, faint ones. This capability enables the discovery of rare phenomena, a census of the variable sky, and tests of fundamental physics at energy scales far beyond those of terrestrial accelerators. However, fully exploiting these opportunities is currently constrained as much by software and data processing methods as by available instrumentation.
The upcoming Vera C. Rubin Observatory's Legacy Survey of Space and Time (LSST) exemplifies the scale of future transient discovery \citep{Ivezic2019}. LSST will scan large swaths of the sky to unprecedented depths. Each potential discovery will be immediately broadcast as an alert, not directly to the entire community but to a set of pre-selected alert brokers tasked with redistributing the data stream to the wider community, while providing easy-to-use tools to search for and visualize astronomical object candidates.
For comparison, the Zwicky Transient Facility (ZTF) \citep{Bellm:19:ZTFScheduler,Graham2018,DeSm2018,Masci2019}, the current survey that the community most commonly uses for transient follow-up, produces $\sim 10^5$--$10^6$ alerts per night, while LSST will produce greater than an order of magnitude more ($\sim 10^7$).

Real-time processing and rapid follow-up of this alert stream is critical for many science cases, with brokering software needing to keep up with the rate of alert creation while maintaining or increasing the number of features it aims to offer to the end user. The alerts emitted by these large surveys include a variety of transient phenomena including, among many other science cases:
\begin{itemize}
    \item Very young Type Ia supernovae (SN\,Ia) with either an early ``flash'' or ``bump'' in their light curves well before the epoch of maximum light (e.g., iPTF14atg; \cite{cao_strong_2015}, SN2017cbv; \cite{Hosseinzadeh_2017} and SN2019yvq; \cite{Miller_2020}).
    \item Luminous fast blue optical transients (LFBOTs) \cite{PrMa2018,Perley2019,HoGo2019,HoPe2020,Perley_2021}, with optical and (sometimes) copious X-ray emission evolving on short timescales.
    \item $\gamma$-ray burst afterglows \cite{NyFr2009,GeMe2012} from either collapsars or neutron star mergers.
    \item Kilonovae associated with binary neutron star mergers \citep{AbEA2017b} such as AT2017gfo \citep{CoFo2017,SmCh2017,kasliwal_2017,AbEA2017f}.
    \item Jetted tidal disruption events whose accretion leads to the launch of a relativistic jet \citep{Bloom2011Sci,Andreoni_2022}.
\end{itemize}
These young and/or fast transient science cases are bolstered by the rise of instruments in other messengers, e.g., Advanced LIGO \citep{aLIGO} and Advanced Virgo \citep{adVirgo} for gravitational waves; e.g., IceCube \citep{AaAc2017} for neutrinos, or other wavelengths; the {\it Neil Gehrels Swift Observatory} mission \citep{GeCh2004}; Fermi's Gamma-ray Burst Monitor (Fermi-GBM) \citep{MeLi2009}; the Space-based multi-band astronomical Variable Objects Monitor (SVOM); and Einstein Probe \citep{Yuan_2022} for $\gamma$-rays and X-rays.

There is a large software ecosystem enabling time-domain astronomy.
For example, the General Coordinates Network (GCN; \citealt{SiRa2023}) and the Scalable Cyberinfrastructure to support Multi-Messenger Astrophysics\footnote{\url{https://scimma.org/}} (SCiMMA) project are platforms where multi-messenger instruments share real-time alerts with the community that can then either be followed up on directly or cross-matched with alert streams.
Depending on the type of transient, it is common for identified objects to be shared with the community on the Transient Name Server\footnote{\url{https://www.wis-tns.org/}} (TNS) or the Minor Planet Center\footnote{\url{https://minorplanetcenter.net/}} (MPC).
These streams are ingested by Target and Observation Managers (TOM), otherwise known as ``marshals,'' which enable coordinated follow-up efforts. Examples include \texttt{GROWTH Marshal} \citep{kasliwal_2019}, \texttt{YSE-PZ} \citep{CoJo2023}, TOM Toolkit \citep{StBo2018}, and \skyportal \citep{WaCr2019,Coughlin_2023}.

Feeding these marshals are the ``enriched'' alert streams from the brokers, including, among others, ALeRCE \citep{FoCa2021}, AMPEL \citep{Nordin:2019kxt}, ANTARES \citep{MaSt2021}, Fink \citep{MoPe2020}, Lasair \citep{SmWi2019}, Pitt-Google, and \babamul (the plans for which we will discuss further below). These brokers filter the optical alert streams to identify targets of interest. For surveys like ZTF and LSST, alerts are produced when a significant ($> 5 \sigma$) residual flux is detected from a point source in a subtracted image, and are distributed via Kafka\footnote{https://kafka.apache.org} in Apache avro format. While each survey provides a different set of data and, therefore, uses a different schema to serialize it, they all include key properties such as:
\begin{itemize}
    \item Position and brightness of the current detection.
    \item Ancillary detection data and higher-level derived values, including real--bogus scores, which helps distinguish real transients from image artifacts.
    \item The associated triplet of science, reference, and subtraction images.
    \item Time-series information about past alert-based detections, non-detections, and forced photometry.
    \item Higher-level metadata about a known astronomical object at the location of the detection (to within some positional uncertainty).
\end{itemize}

To identify the most interesting objects for particular science cases, brokers can cross-match alerts against static catalogs (e.g., Gaia DR3, \cite{Gaia2023}; PanSTARRs DR1, \cite{ChMa2016}; milliquas, \cite{Flesch_2023}; NED LVS, \cite{Cook_2023}) and look for specific properties using machine learning classification pipelines (e.g., AstroM3, \cite{rizhko2024astrom}; BTSbot, \cite{Rehemtulla+2024}; ACAI, \cite{duev2021phenomenologicalclassificationzwickytransient}; Maven, \cite{zhang2024mavenmultimodalfoundationmodel}). All brokers provide their own filtering system that makes use of the ``enriched'' alert data to identify objects of interest for specific science cases. Depending on the broker, the filters can be from a standard set based on community feedback or customized by the user directly. Furthermore, they may run in real-time as alerts are coming in, or anytime after the alert data is processed to perform archival searches. So far, alert brokers have been providing these services predominantly for the ZTF alert stream. However, now that a number of surveys providing real-time alert streams will overlap in space and time, we---as a community---have an opportunity to enrich each survey with the data products from the others.

Specifically, to supplement the LSST alert stream, a variety of other optical systems such as the La Silla Schmidt Southern Survey (LS4; \citealt{miller2025lasillaschmidtsouthern}) and ZTF will be trailing the LSST footprint daily. To maximize the science synergies enabled by these coordinated observations, brokers will need to perform joint filtering on these alert streams.
This will be essential to readily identify, for example, fast transients within these streams.
The relatively slow cadence of LSST and other optical surveys' limited depth make such identification difficult. Thus, the rate of evolution of these phenomena cannot be precisely measured when alerts from these surveys' streams are used separately.

It is with these considerations in mind that we present \texttt{BOOM}, an astronomical alert broker that builds upon our experience with \kowalski\footnote{\url{https://github.com/skyportal/Kowalski}}, an open source, multi-survey data archive and alert broker \citep{DuMa2019}. \kowalski has been used in production by the ZTF collaboration for over seven years, with \skyportal as its ``marshal``.
In this paper, we will describe several important developments and design choices made with \boom that ready it for the upcoming LSST era.
Although this paper focuses mainly on the design and conceptual framework of \texttt{BOOM}, we encourage the interested reader to explore the repository alongside this text \footnote{https://github.com/boom-astro/boom}, and read its documentation.

As an example of important design choices,
since multiple observatories will observe the LSST footprint concurrently,
\boom has as its top priority,
the ability to jointly filter on multiple alert streams,
a relatively unique capability in the broker community.
Furthermore, one of the main differences between
\boom and \kowalski---and most other alert brokers---is moving away from
Python and instead writing the project in \texttt{Rust},
a compiled programming language with performance characteristics that make
it much better suited to attaining a high throughput of alerts at scale.
Processing additional alert streams in parallel increases the data volume
by at least an order of magnitude,
and jointly filtering on multiple streams requires even more computation,
as every alert packet needs to be cross-matched with all other
overlapping surveys' alert streams.
While not relying on \texttt{Python}
may have the downside of making contributions from astronomers---who
typically write in Python---more challenging,
our experience running \kowalski in production has shown us that requiring
computer science knowledge to make use of an alert broker will significantly
limit its use, regardless of complexity. In the following sections,
we will describe the filter building tool we have built to ameliorate this
challenge, where the user is provided with a no-code alternative to write
complex queries.

In this paper, we describe the components built within \texttt{BOOM},
focusing on design choices useful in preparation for LSST data scales.
We describe \texttt{BOOM}'s key features in Sec.~\ref{sec:pipeline}.
We demonstrate some of \texttt{BOOM}'s capabilities, including throughput measurements, in Sec.~\ref{sec:science}.
We describe the future of \boom and how we envision its role in the community in Sec.~\ref{sec:conclusion}.

\section{\boom Framework}
\label{sec:pipeline}

\begin{figure*}[t]
    \centering
    \includegraphics[width=7.2in]{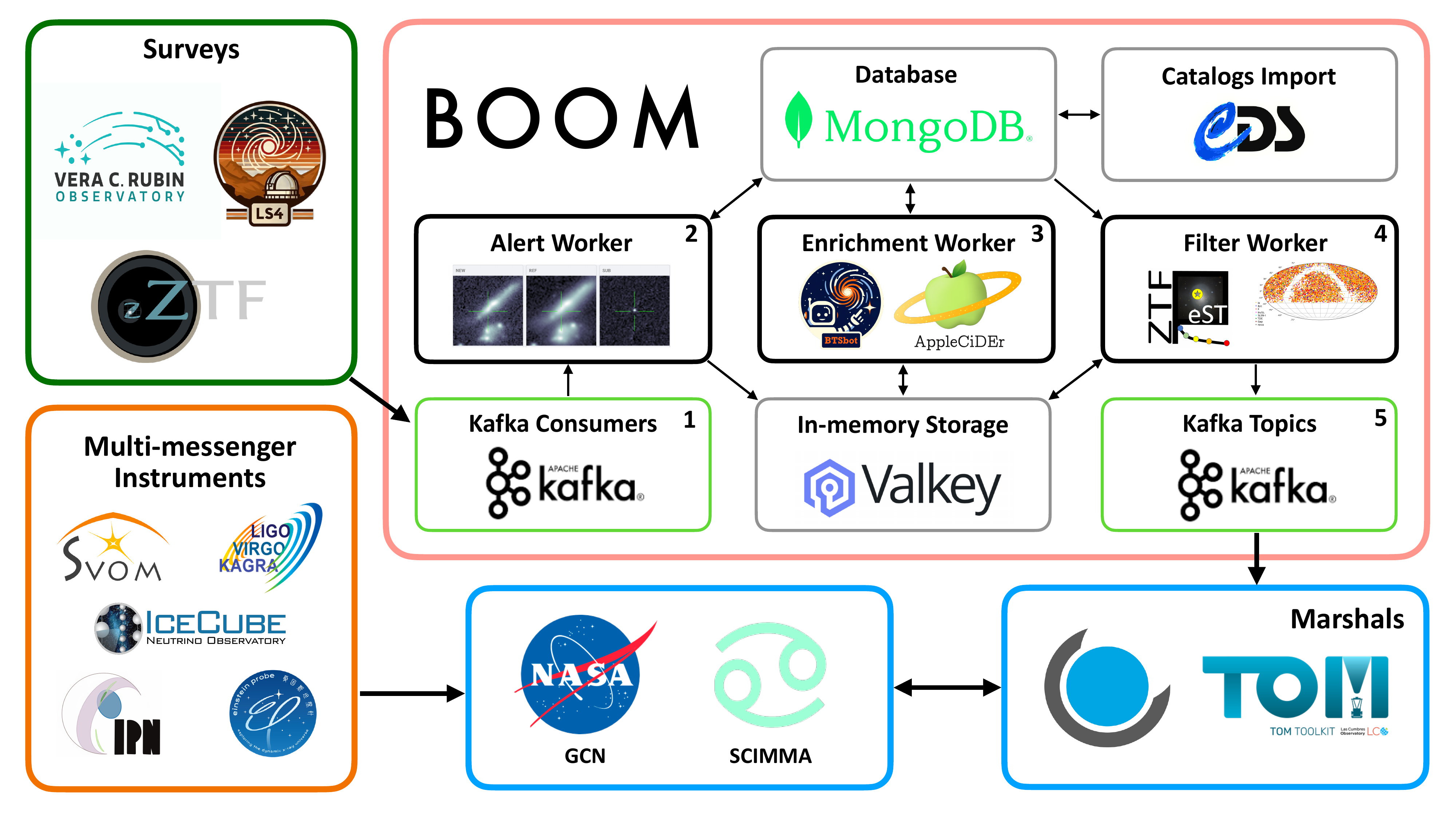}
    \caption{Flowchart for BOOM.}
    \label{fig:flowchart}
\end{figure*}

In this section, we present the key design features and implemented capabilities within \boom.
\boom is designed for full parallelization. The database and alert processors all scale horizontally, allowing additional workers to be added at any stage to accommodate changes in workload. Alerts can be processed in any order, meaning they do not need to follow a strict time sequence.
\boom operates with workers, separating the machine learning, cross-matching, filtering, ingestion, etc. into different processes. Each of these workers are described herein.

\subsection{Input/Output through \texttt{Apache Kafka}}

\texttt{Apache Kafka} has become the gold standard for astronomical alert brokering due to its scalability, fault tolerance, and capacity to handle large data volumes for a wide variety of production-grade software in academia and industry. Optical alerts from surveys like ZTF and LSST are transmitted as Avro packets\footnote{\url{https://avro.apache.org}} over \kafka, which means that the ability to feed from \kafka topics---which represent one data stream that a client can read from---is absolutely required by any alert brokering software. Its ecosystem of libraries, available for all major programming languages, makes it extremely easy for the client to develop pipelines around it. For these reasons, adopting \kafka as our downstream data sharing system ensures compatibility with existing downstream services, which are also designed to consume alerts from \kafka . For each survey supported by our software, an associated \kafka consumer has been developed to feed from the \texttt{Avro}-formatted alerts. The consumers take advantage of \kafka's partitioning feature (where one topic is broken down in N partitions that a client can read from independently, with multiple processes) to process any given survey's alert stream(s) in parallel, maximizing the input rate. \texttt{BOOM}'s output, as described in subsection \ref{sec:filter}, is also serialized to \texttt{Avro} and produced to a \kafka cluster.

\subsection{Job scheduling with \valkey}

Whereas \kafka shines when it comes to reliable message sharing at scale across the Web, it is not the most performant solution for interprocess communication as its topics are stored on disk, which results in throughput limited by the host's I/O capabilities.
Instead, we opted for \valkey, a high-performance and open source in-memory datastore backed by the Linux Foundation.
\valkey can be used for a variety of workflows, including caching and message queues.
Unlike \kafka, all data stored by \valkey live in RAM, ensuring much higher throughput than physically possible with data stored on HDDs or SSDs. However, a `persistence` feature can also be enabled, which allows \valkey to periodically create backups on disk, so that its content can be restored in the event of a catastrophic failure. This process is performed asynchronously and did not show a noticeable impact on performance.
Once data is read from the various survey's \kafka topics, \texttt{BOOM}'s \kafka consumer stores alerts to be processed in \valkey lists; these are simple array data structures from which processes on the same machine or network can read from concurrently, one element at a time or in batches. When read from a list, messages are removed from it and one message can only be retrieved by one process. This is precisely what we need for \texttt{BOOM}, where alert processing is not performed by one but by many processes, to parallelize over the data streams. Message queues used by \texttt{BOOM}'s workers to communicate with each other only share a minimal amount of information: mostly pointers to database documents, and unique identifiers of alert packets. Thus, they do not have a significant memory impact. However, memory usage remains a concern when relying heavily on in-memory storage technologies for larger data products, such as the original \texttt{avro} alerts at the first stage of processing. With this in mind, \boom sets limits on how many alert packets are stored in memory at once. Moreover, since \valkey is only used as a job queue, all lists are meant to be temporary and consumed as they are filled. As long as \boom is configured to handle incoming alerts with little to no throttling, which means providing it with sufficient compute capabilities, these lists remain fairly empty and so does the overall memory usage.

\subsection{Spatial query-ready database with \mongo}

\mongo has proven to be a highly effective choice for alert brokering, as demonstrated by its successful implementation in \kowalski. Its cross-language support, flexibility, and powerful query language make it well-suited for building complex filtering pipelines for transient alerts. Namely, the \texttt{aggregation pipeline} feature has allowed us to define not only complex queries but also pipelines of queries with multiple stages,
such as a cycle of filtering
(\texttt{\$match} stage),
computing (\texttt{\$project}, \texttt{\$addField}, \texttt{\$lookup} ... stages) steps,
well-suited to implement astronomical alert filtering pipelines. \mongo offers both performance and scalability, essential for handling large data volumes efficiently. Its built-in compression also simplifies data storing requirements.
While \texttt{PostgreSQL} was a potential alternative,
it would have required schema enforcement, which in turns requires database migrations whenever a new astronomical catalog is integrated for cross-matching or a new survey's support is added.
Additionally, \mongo natively supports \texttt{GeoJSON} indexes for fast spatial queries, such as cone-searches or nearest neighbor searches without any client-side implementation or extensions required, features that \texttt{PostgreSQL} does not implement natively.
Just like \kowalski, \boom relies very heavily on \mongo's native support for cone-searches between alert streams and archival/static catalogs, and on it's \texttt{aggregation pipeline} feature to design and execute complex user-defined filters. When it comes to our data model, alert packets are dividing into 3 distinct collections (\texttt{MongoDB's} equivalent of a table, as found in a relational context):
\begin{itemize}
    \item The \texttt{Alert} collection, containing an alert's \texttt{candidate}, metadata about the latest detection that resulted in the alert being sent. To these we later append time-dependent data products, such as machine learning scores and ``pre-computed'' features to facilitate the implementation of user-defined filter, as described in subsection \ref{sec:enrichment}. Entries of the collection are indexed on the alert's candidate ID (\textit{candid}, a unique identifier provided by the associated survey), its object ID (\textit{objectId}, an identifier for this astronomical object, most often purely position based: detections made at the position of a previous alert will be attributed the same \textit{objectId}), and its position.
    \item The {Object} collection, containing lightcurve data products (concatenated from the time-limited lightcurves provided by each alert for the same object, as surveys provide only N days worth of past detections), and matches with other catalogs and surveys. For archival catalogs matches all the relevant metadata is stored in this collection, whereas for alert-based survey matches only the survey's \textit{objectIds} are stored to enable lookups when user-defined filters are run. Entries of this collection are indexes on \textit{objectId}, and on the object's position (taken from its first alert ingested by \texttt{BOOM}, which is subject to change as a flux-averaged centroid may be more adequate).
    \item The {Cutout} collection, simply containing the science, reference, and difference image cutouts from the alert packet. This collection is also indexed on the alert's \textit{candid}, to enable quick lookups of alert images based on their identifier.
\end{itemize}

With this data model, the data-heavy images that cannot be queried like other data products would are stored on their own and can be optionally retrieved alongside alerts using lookups, and object-specific data products (i.e. positional based) such as cross-matches and lightcurves are stored in one place instead of on every alert (as served over \kafka by the various surveys), which would yield considerable duplication and increase data storage requirements.

\subsection{Parallelized and Distributed Alert Processing}

\boom employs a different architectural approach than its predecessor \kowalski, using dedicated worker types for each processing stage rather than a single monolithic worker design. In \kowalski, alert processing was parallelized using a cluster of identical workers, where each worker was responsible for the complete end-to-end processing of individual alert packets. This included performing database insertions and queries such as cross-matching with archival catalogs, running user-defined filters, inserting and updating alerts and objects, and running machine learning models. While this design simplified deployment and management, it suffered from significant inefficiencies that prevented it from scaling up sufficiently and efficiently.

The single worker-type approach creates several unavoidable bottlenecks. First, forcing a sequential processing of alerts one at a time prevents the system from taking advantage of batch operations that are essential for both database efficiency and machine learning performance. Database queries such as retrieving or inserting documents benefit substantially from being performed over batches of entries rather than one by one, as this reduces network round-trips and the overhead incurred by each operation. Similarly, machine learning models are designed to parallelize inference over multiple inputs simultaneously (to leverage hardware acceleration such as GPUs, though it may improve performance in a CPU-based environment) rather than processing them sequentially.

Perhaps more critically, the monolithic worker design creates inflexible scaling constraints. When a single operation represents a significant portion of processing time, in single end-to-end worker scenario, the only available solution is to add more copies of the same one worker, inevitably scaling all of its operations regardless of whether they constitute bottlenecks. In \kowalski's case, processing time was dominated primarily by user-defined filters and secondarily by machine learning inference, yet scaling these bottlenecks required also scaling other processing steps that were not performance-limiting factors, in turn unnecessarily scaling database load, CPU usage, and memory consumption, slowing down the overall system.

\boom addresses these limitations through a multi-tier architecture with dedicated worker types for ingestion, inference, and filtering operations. While initial alert ingestion remains sequential, both inference and filtering are handled by specialized worker types that process batches of alerts, dramatically reducing the number of database operations required. This architectural separation enables independent scaling of different processing stages, allowing administrators to increase compute resources only where bottlenecks occur. This resource optimization becomes particularly important given \texttt{BOOM}'s expanded multi-survey capabilities, which introduce numerous additional database operations compared to single-survey systems. This results in more efficient hardware utilization and lower overall resource consumption, while providing superior processing throughput and flexibility.

An alternative approach might consider using single end-to-end workers that process not one but batches of alerts through sequential processing stages with parallelization where possible. However, this design creates a fundamental latency bottleneck: since certain initial operations like deserializing \texttt{Avro} packets and cross-matching with archival catalogs and surveys must be performed sequentially on individual alerts, a worker cannot begin machine learning inference on a batch until it completes all preliminary processing for every alert. As batch sizes increase to improve machine learning efficiency, latency from an alert being emitted and it being processed increases proportionally because the worker must sequentially process all alerts in a batch before any can proceed to inference, and then filtering. The only way to reduce this latency would be to decrease batch sizes and add more workers, but this ultimately converges to the inefficient single-worker-single-alert scenario, negating the benefits of batch processing entirely.

\begin{figure*}[t]
  \centering
  \resizebox{\textwidth}{!}{% \documentclass[tikz,border=10pt]{standalone}
% \usetikzlibrary{positioning, shapes.geometric, arrows.meta, fit, calc}
% \begin{document}
\begin{tikzpicture}[
    font=\sffamily,
    node distance=12pt,
    process/.style={rectangle, rounded corners=5pt, draw=black!70, fill=white,
        very thick, align=center, text width=6cm, minimum height=0.8cm},
    process small/.style={rectangle, rounded corners=5pt, draw=black!70, fill=white,
        very thick, align=center, text width=5cm, minimum height=0.8cm},
    process very small/.style={rectangle, rounded corners=5pt, draw=black!70, fill=white,
        very thick, align=center, minimum height=0.8cm},  % No fixed text width
    decision/.style={diamond, aspect=2.2, draw=black!70, thick, fill=white,
        align=center, inner sep=1pt, text width=3.6cm},
    flowarrow/.style={-latex, thick},
    box/.style={rounded corners=10pt, inner sep=10pt, draw=none,
        fill opacity=0.5, text opacity=1},
    title/.style={font=\bfseries\large, align=center, text opacity=1},
]
% --------------------------------------------------
% ==== ALERT WORKER (LEFT COLUMN) ====
% --------------------------------------------------
\node[box, fill=red!20, fit={(-12, -2.8) (-1, 12)}] (alertbox) {};
\node[title] at (-6.5,11.6) {Alert Worker};
% \node[font=\bfseries] at (-3.4,-0.5) {1};
\node[process] (a1) at (-6.5,10.6) {In-Memory Storage};
\node[process, below=of a1] (a2) {Retrieve Alert Avro Packet};
% Main "no" branch continues downward
\node[process, below=of a2] (a5) {Insert Alert \& Cutouts};
% \node[process, below=of a4] (a5) {Insert Cutouts};
\node[process, below=of a5] (a6) {Xmatch w/ Other Surveys' Objects};
\node[decision, below=of a6] (a7) {Object Already Exists?};
% Short "yes" branch to the left
% \node[process, below left=of a7] (a8yes) {Update Object Lightcurves \\ + Survey Xmatches};
\node[process small, below left=1.2cm and -1cm of a7] (a8yes) {Update Object's \\ Lightcurves \\ + Survey Xmatches};
% Main "no" branch continues downward
\node[process small, below right=0.5cm and -1cm of a7] (a8no) {Xmatch w/ Archival Catalogs};
\node[process small, below=of a8no] (a9) {Insert Object w/ Lightcurves \\ + Survey Xmatches \\ + Archival Xmatches};
\node[process, below=4.4cm of a7] (a10) {Candidate IDs};
\node[process, below=of a10] (a11) {In-Memory Storage};
% Arrows
\draw[flowarrow] (a1)--(a2);
\draw[flowarrow] (a2)--(a5);
\draw[flowarrow] (a5)--(a6);
\draw[flowarrow] (a6)--(a7);
\draw[flowarrow] (a7.south west)--node[above left, font=\small]{Yes}(a8yes);
\draw[flowarrow] (a7.south east)--node[above right, font=\small]{No}(a8no);
\draw[flowarrow] (a8yes)--(a10);
\draw[flowarrow] (a8no)--(a9);
\draw[flowarrow] (a9)--(a10);
\draw[flowarrow] (a10)--(a11);
% --------------------------------------------------
% ==== ML WORKER (CENTER COLUMN) ====
% --------------------------------------------------
\node[box, fill=blue!20, fit={(0.2, -2.8) (6.8, 12)}] (mlbox) {};
\node[title] at (3.5,11.6) {Enrichment Worker};
% \node[font=\bfseries] at (5.7,-0.5) {2};
\node[process] (m1) at (3.5,10.6) {In-Memory Storage};
\node[process, below=of m1] (m2) {Retrieve Candidate IDs};
\node[process, below=of m2] (m3) {Fetch Alerts w/ Lightcurves \\ + Survey Xmatches \\ +  Archival Xmatches};
\node[process, below=of m3] (m4) {Compute Metadata \& Lightcurves alert features};
\node[process, below=of m4] (m5) {Prepare Metadata, Lightcurves, Image ML inputs};
\node[process, below=of m5] (m6) {Run Inference for Each Model};
\node[process, below=of m6] (m7) {Update Alerts w/ features and ML scores};
\node[process, below=of m7] (m8) {Candidate IDs};
\node[process, below=of m8] (m9) {In-Memory Storage};
\foreach \i/\j in {m1/m2,m2/m3,m3/m4,m4/m5,m5/m6,m6/m7,m7/m8,m8/m9}
  \draw[flowarrow] (\i)--(\j);
% --------------------------------------------------
% ==== FILTER WORKER (RIGHT COLUMN) ====
% --------------------------------------------------
\node[box, fill=yellow!30, fit={(8, -2.8) (15, 12)}] (filtbox) {};
\node[title] at (11.5,11.6) {Filter Worker};
% \node[font=\bfseries] at (14.7,-0.5) {3};
\node[process] (f1) at (11.5,10.6) {In-Memory Storage};
\node[process, below=of f1] (f2) {Retrieve Candidate IDs};
\node[decision, below=of f2] (f3) {Passes Any Filter?};
% Short "no" branch to left
% \node[process very small, left=2cm of f3] (f4no) {Skip};
\node[process very small, below left=0.6cm of f3] (f4no) {Skip};
% Main "yes" branch continues downward
\node[process, below=0.8cm of f3] (f4yes) {Fetch Alert w/ metadata \\ + Lightcurves \\ + Images \\ + Survey Xmatches \\ + Archival Xmatches};
\node[process, below=of f4yes] (f5) {Combine Alert data + Passed filters annotations};
\node[process, below=of f5] (f6) {Serialize to Avro Packet};
\node[process, below=of f6] (f7) {Kafka Topic};
% Arrows
\draw[flowarrow] (f1)--(f2);
\draw[flowarrow] (f2)--(f3);
\draw[flowarrow] (f3)--node[above left]{No}(f4no);
\draw[flowarrow] (f3)--node[right]{Yes}(f4yes);
\draw[flowarrow] (f4yes)--(f5);
\draw[flowarrow] (f5)--(f6);
\draw[flowarrow] (f6)--(f7);
\end{tikzpicture}
% \end{document}
}
  \caption{Decision tree workflow of each boom worker}
  \label{fig:workflow}
\end{figure*}
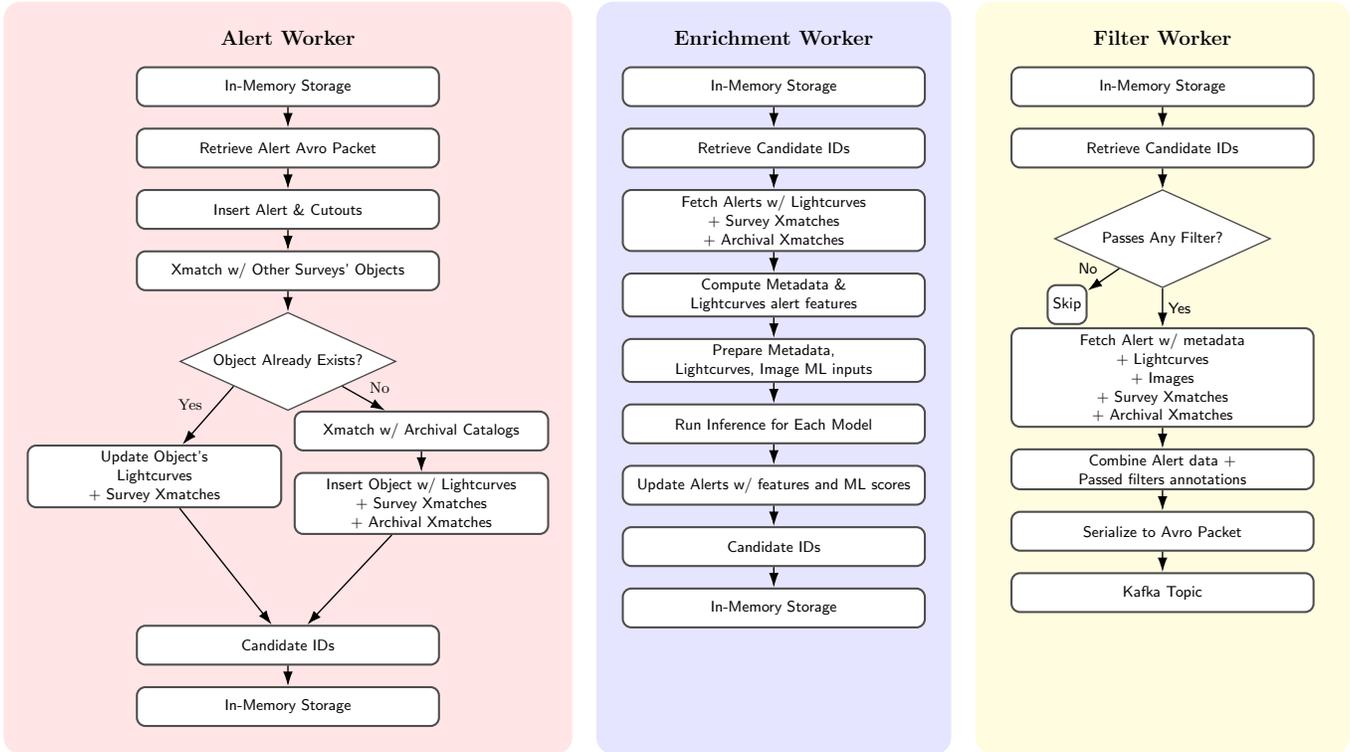

Next, we will describe exactly how the alert processing steps mentioned above have been split into multiple ``workers'', as illustrated in Fig.~\ref{fig:workflow}.

\subsubsection{Alert Ingestion Worker}

\texttt{BOOM}'s first worker type is the \texttt{Alert Ingestion} worker. It feeds from the \valkey queue that has been populated with \texttt{Avro} alert packets by the \kafka consumer(s), and its main role is to ingest crossmatch-enriched reformatted alerts to the \mongo collections of a given survey. It processes one alert packet at a time (multiple workers of this type are spawned to handle the load and maximize parallelization), going through the following steps:
\begin{itemize}
\item Read the \texttt{Apache Avro} byte data to deserialize into \texttt{Rust} \texttt{struct}s matching the schema provided by the survey emitting the alerts. Here, we rely heavily on the \texttt{serde} and \texttt{apache-avro} crates; the former also allows us to customize the deserialization logic to modify the alert schema of each survey, in an effort to reduce some of the differences between different surveys schemas, and to address some of their inefficiencies.
\item We separate the candidate metadata about the current detection, \textit{candid}, and \textit{objectId} from its cutout images and time-series.
\item The candidate, \textit{candid}, and \textit{objectId} are stored in the alert collection.
\item The science, reference, difference cutout images are stored in the cutout collection.
\item For new objects---identified as the objects for which no database entry exists in the object collection---we cross-match the candidate's position with a number of static/archival catalogs. Since a new \textit{objectId} is only generated when we receive the first ever alert at a given right ascension and declination ($\pm$ some uncertainty that varies based on a survey's hardware and alert pipeline), cross-matches with static catalogs only need to happen once for a given \textit{objectId}. Indeed, as the input position for an object and archival catalogs do not change over time, the results of cross-matches do not change either. Solar system objects are the only exception to this rule since their position is ever changing, but cross-matches with static catalogs are not relevant for these to begin with. Here, the radius used is the maximum between the positional uncertainty of the alert survey and the positional uncertainty of the instrument that was used to build the static catalog. This value can be configured.
\item Similarly, we cross-match every new alert with the object collection of other surveys supported by \boom. While the position is still immutable for a given \textit{objectId}, here the catalogs we cross-match against are dynamic, populated with new objects as the surveys generate alerts at new positions. Therefore, these cross-matches happen for every new alert and not only new objects. The cross-match radius is also defined as the maximum positional uncertainty between the two cross-matched surveys. The cross-match information resulting from this operation is a simply document with the names of the survey's we cross-match against as keys, and an array of matching object IDs from these surveys. This survey-matches information may be used by both enrichment and filter workers, to create joint survey lightcurves on-the-fly, where required.
\item Last but not least, we create or update the object collection for the alert's survey. New objects get a new entry (indexed by their respective object ID) containing time-series data products (lightcurves of previous candidates, non-detections, and forced-photometry), cross-matches with archival catalogs, and cross-matches with other supported alert surveys. Existing objects have their associated database entry updated, with new data points appended to their time-series data products, updated survey-matches always kept up-to-date.
\end{itemize}

Once an alert has been ingested, its \textit{candid} is pushed to another \valkey queue for the next worker type to read from: the \texttt{Enrichment} worker.

\subsubsection{Enrichment Worker}
\label{sec:enrichment}

While we have not observed any obvious advantage to ingesting alerts in bulk, there is a clear advantage to running machine learning models over batches of inputs, as these can easily be parallelized by the various machine learning frameworks available to us using hardware accelerators (e.g. GPUs, TPUs). So, the Enrichment worker will not read and process only one \textit{candid} at a time, but a batch with a maximum size, e.g. 1000. We use an aggregation pipeline to retrieve the full batch of candidates at once, with full light curve(s) and cutouts from the database. These are then converted into the expected format for each machine learning model \boom supports. If multiple models expect the same input features, these are only computed once to avoid unnecessary work. At this time, \boom runs all 5 ACAI classifiers \citep{Duev+2021}, and BTSbot \citep{Rehemtulla+2024}. These models have already been used in Kowalski successfully for a number of programs, including fully automated follow-up, e.g. BTSbot.
Although most popular machine learning frameworks have been designed in Python and, therefore, not usable as in other languages, there are a number of solutions to port \texttt{Python}-trained models over to a \texttt{Rust}-based pipeline. We explored the following:

\begin{itemize}
    \item If the Python ML framework used is built around a C-based low-level
        library, bindings are often available to run the same models in Rust
        (e.g. TensorFlow, PyTorch). These options proved to lack community
        support and documentation, making it an unsustainable approach.
    \item As we are relying on a ``compartmentalized'' architecture with a
        worker type dedicated to machine learning, one could simply implement this worker directly in
        \texttt{Python}. In fact, \texttt{BOOM}'s architecture originally had
        been designed to allow for inter-operability between languages. However, we
        later decided to focus on a \texttt{Rust}-only approach as we
        successfully converted \kowalski's \texttt{Python}-trained model
        to a format suitable for running directly in \texttt{Rust} (described
        below).
    \item The \texttt{pyo3} crate allows for seemless integration of Python code
        in a \texttt{Rust} runtime. Using this package, one can directly run
        Python code within a rust program. However, this did not prove to yield
        a significance performance improvement compared to the complexity added
        to the software. In addition, a lack of documentation and
        community-backed examples applied to machine learning steered us away from this option.
    \item Last but not least, most standard machine learning
        implementations---regardless of the framework used---can be converted to
        an open-source framework and language-agnostic format called \texttt{ONNX} (Open
        Neural Network eXchange) \footnote{\url{https://onnx.ai/}}.
        \texttt{ONNX} defines a set of common
        operators used to represent models trained with most ML frameworks in a
        graph-like format. Community-driven Python packages for both PyTorch and
        TensorFlow enable conversion of trained model to the
        \texttt{ONNX} format, which can then be loaded into any language
        with an \texttt{ONNX} runtime (which is, most). In
        Rust, we used the \texttt{ort} crate (https://ort.pyke.io/). Also, \texttt{ONNX}'s graph optimizer is able to removex
        unnecessary, redundant, or suboptimal operators and nodes from its graph-representation, sometimes resulting in faster inference than possible in the framework used to train the converted model.
\end{itemize}

After experimenting with the four approaches, integrating \texttt{ort} in a \texttt{Rust}-based program to run \texttt{Python}-trained models converted to \texttt{ONNX} yielded the best ``performance vs. complexity'' ratio.

So far, all \kowalski models have been converted to \texttt{ONNX} and have been implemented in \boom. Furthermore, new models and architecture are being developed to tackle LSST-era challenges, such as \texttt{AppleCiDEr} \cite{junell2025applyingmultimodallearningclassify,xu2025applecideriispectranet}. Implemented and trained in PyTorch before being converted to \texttt{ONNX} and integrated with \texttt{BOOM}, \texttt{AppleCiDEr} is a multimodal machine learning based framework for early transient classification that combines four complementary data modalities: photometry, image cutouts, metadata, and---optionally---spectra. It utilizes transformer encoders for light curves, a multimodal convolutional neural network (CNN) with domain-specific towers for images and metadata, and a dedicated CNN for spectral data. Trained on real ZTF alerts, \texttt{AppleCiDEr} achieves high accuracy across diverse transient classes. Since spectral data are not available in real time, only the photometry and image–metadata models are currently integrated into \boom.

While maintaining \kowalski, we identified the following:
\begin{itemize}
    \item A multitude of identical features that most user-defined filters relied on, computed from alert metadata and lightcurves (e.g. is this a potential asteroid, a star, near a brightstar, ...). This meant that many filters were computing the same values over and over again. This appeared to be a clear waste of database compute resources and time.
    \item A number of transient identification pipelines which relied on \kowalski's API to periodically query for new alerts with minimal filtering, rather than the built-in user-defined filter system. These then performed more complex computation (such as determining the peak of a large lightcurve in each band, and/or evaluating the rate of evolution before and after peak) which required too much database-specific knowledge to be implemented with \mongo operators as used by the user-defined filters.
\end{itemize}

To address both of these issues, the \texttt{Enrichment} worker now computes a number of these features directly. These depend on the data products available and therefore on the survey of origin. The features currently implement are subject to change, and new ``pre-computed'' features will be added to BOOM.
These are computed in the same loop responsible of generating ML model inputs. They can be used both in user-defined filters, and while performing archival searches through \texttt{BOOM}'s RESTful API \citep{rest}. With multi-survey support in mind, we aim to add additional lightcurve-based features that not only rely on the current object's lightcurve, but concatenated with those from other matching surveys.

Just as alert data products are retrieved from the database in batches, we use \mongo's batch update operator to update all processed alerts at once with ML scores and features. Once the batch update is finished, the \textit{candids} of processed alerts are sent back to a \valkey queue for the next and last worker type to read from: the \texttt{Filter} worker.

To further improve throughput, we are developing an optional GPU worker that offloads batch inference to a hardware accelerator. By batching alert processing the BTSBot classifier on $\sim$1000 alerts simultaneously on a single NVIDIA P100, we observe an approximately ten-fold improvement in enrichment throughput compared to CPU-only inference. Building on this GPU-accelerated architecture, we are developing a lightcurve fitting library that performs both nonparametric and parametric model fitting directly within \boom's \texttt{Rust}-based pipeline. For nonparametric fitting, Gaussian Process (GP) regression is applied per-band and jointly across bands (using a two-dimensional time--wavelength kernel) to produce smooth interpolations of sparsely sampled lightcurves, from which features such as rise and decay timescales, peak magnitudes, and color evolution are extracted. For parametric fitting, a particle swarm optimization algorithm performs model selection across a library of analytical lightcurve models \citep{deSoto_2024}, with posterior uncertainties estimated via stochastic variational inference (SVI). All fitting stages, including GP hyperparameter selection, PSO model fitting, and SVI, are implemented as batch CUDA kernels, enabling thousands of sources to be processed in parallel on a single GPU. The impact of these GPU-accelerated algorithms on both system throughput and transient classification performance will be discussed in detail in a future publication.

\subsubsection{Filter Worker}
\label{sec:filter}
\texttt{BOOM}'s filter worker is the last worker type to process alerts before producing an output that other systems can read from. At initialization, the filter worker loads user-defined filters from the database. These filters are defined as MongoDB aggregation pipeline running on the alert collection, composed of a succession of \$project and \$match stage to transform and filter on the alert data iteratively. User-defined filters---as  written by BOOM's users---assume that all data products are available for them to filter on. However, since these data products are divided into different collections in our data model, we pre-pend all user-defined pipelines with additional lookup stages (e.g. retrieving various lighcurves and cross-match information from the object collection). Instead of pre-pending all user defined filters with the same ``lookup'' stages that retrieve all available data products, we scan each user-defined filter to identify which ones they make use of, so we can decide where in their pipeline to add which lookup operations. This ensures that no unnecessary computation is performed. However, this system is obviously dependent on the order of the operations performed by user-defined filters. If a filter uses data products found in the object collection early on, lookups will always be performed first and for all the alerts that are filtered. If on the other hand, the filters first use candidate metadata, pre-computed features, and ML scores before potentially using object-level data products, lookups will only run for the small subset of alerts that pass the first filtering stages. User-defined filters are also encouraged to define an `annotations` key in their output document with a final \$project stage. This key may contain any field of interest for this particular filter, that they would like to see in \texttt{BOOM}'s output.

The \textit{candids} sent by the Enrichment worker to the Filter worker are read and filtered on in batches. Rather than filter sequentially on one alert at a time, each filter runs in turn on all alerts at once. This dramatically saves on DB usage and reduces the time required to filter on alerts. Once all filters have run on a batch of \textit{candids}, we are left with a hashmap where the keys are \textit{candids} that have passed at least one filter, and the values are the list of filters ids that each \textit{candid} has passed, and annotations if any. Then, we query the database to retrieve all the relevant data products for the subset of alerts that have passed at least one filter, and build \texttt{BOOM}'s final output: the \texttt{Alert} struct. This struct is identical for all surveys, but populated with a custom logic for each. It contains:
\begin{itemize}
    \item metadata about the object and alert, such as IDs,
        position, and survey of origin.
    \item a list of `Classifications`, defined by the classifier
        name and score.
    \item a list of `Photometry`, defined by their time, band,
        flux data, and pipeline of origin (alert vs forced
        photometry). This only contains photometry from the alert's own survey, not from other survey's it matched with.
    \item the 3 cutouts: science, reference, difference
    \item the list of filters they passed, defined by their
        IDs and an optional `annotations` field.
    \item the list of `archival-matches`, containing all cross-
        matches with archival catalogs, as performed by the
        `Alert` worker. Each cross-match entry is characterized
        by its catalog name, and all the fields that are relevant
        for this catalog.
    \item a map of `survey-matches`, containing all the cross-
        matches with other surveys processed by `BOOM`. We use the survey names as keys, and values use the same schema as the \texttt{Alert}, but of course without \texttt{survey-matches}. A downstream user may build a multi-survey lightcurve by concatenating the alert's photometry array with those found in its survey-matches.
\end{itemize}

This schema is subject to change and is expected to evolve as the first instances of \boom are deployed to production, with downstream systems connected to their respective outputs.

As mentioned in the introduction, \texttt{BOOM}'s main concern is enabling multi-survey filtering. Since alerts from one survey are matched with objects from all other supported surveys and the matching surveys' \textit{objectIds} have been stored in the object collection, user-defined filters can make use of other survey's lightcurves. This is made possible by the addition of lookups in the user-defined pipelines, to other survey's object collection using the \textit{objectIds} stored in the database. Thereafter, users' filters can concatenate these lightcurves and use them as one, or simply make use of the matching information. Section \ref{science_validation} showcases what these features enabled during a joint-stream experiment conducted in May 2025, and section \ref{production} demonstrates its application to the public LSST alert stream released on Feb 2026.

\subsection{A RESTful API for archival searches}

While the elements of \boom described so far have been built with real-time operations in mind, since all data products are stored in its database, these can be queried after the fact.
Just like \kowalski exposed an HTTP RESTful API to let its users perform bulk archival searches or to run semi-real-time pipelines,
\boom is deployed alongside a similar API. API users can query any alert collections, as well as any other archival catalogs used by \boom during real-time processing for cross-matching. For advanced users, the API directly exposes \mongo's various query features (e.g., find and aggregate queries).
With the user-defined filters defined as aggregation pipelines, these can by design run on large batches of data and not only a set of candidates. This resulted in the implementation of filter ``re-running'' features, where users of the API can re-run their filters over entire night's worth of alerts, or alerts of known objects. Such a feature can be used while designing user-defined filters, and to validate their results before enabling them in production. Moreover, API endpoints have been implemented to return all data products associated to a given survey's \textit{objectId}, including data products from other matching surveys.

\section{Deployments, Science Validation and First Results}
\label{sec:science}

\subsection{Throughput Testing}

To ensure that \boom is able to handle the additional load from the
Rubin alert stream ($\sim 10^4$ alerts every 30\,s) and beyond,
a throughput test was performed over varying
numbers of worker processes.
One night of ZTF alerts (2025-03-11) was ingested,
cross-matched against the NED LVS catalog,
enriched with the ML models scores and features listed in section \ref{sec:enrichment},
and finally filtered against 25 representative filters.
\kowalski was also run for the same scenario with varying worker
process counts for comparison.
For both brokers, the database is pre-loaded with the full object-level data
products (lightcurves, cross-matches) for all objects that have been detected
prior to 2025-03-11. This ensures that the throughput test is representative
of what is expected in production, where most alerts are emitted for
existing---and often persistent---objects that may have thousands of data
points in their light curves. The selected night consists of 29,142 alerts total, observed across 105 exposures for a total of 70 unique ZTF fields. Since the database is pre-loaded with all previous history for the objects associated to the alerts being processed (most of which had been detected already: 27,948, or $\sim 95.9 \%$), processing a larger or smaller batch of alerts will not have any impact on the measured throughput, and only increase the time required to explore the worker count parameter grid.
The machine used for throughput testing has a 2.9 GHz AMD EPYC 7002 processor
with 32 cores, 64 threads, and 128 GB of 2933 MHz DDR4 memory.
Data were written to a 12 Gb/s 7200 RPM hard drive
(effective hard drive write and read speeds are much lower, no more than 250 MB/s),
and ML inference was performed without GPU acceleration.
The code and datasets to reproduce the results are available
from~\cite{JegouDuLaz2026BoomCalkit}.

\begin{figure}
    \centering
    \includegraphics[width=3.5in]{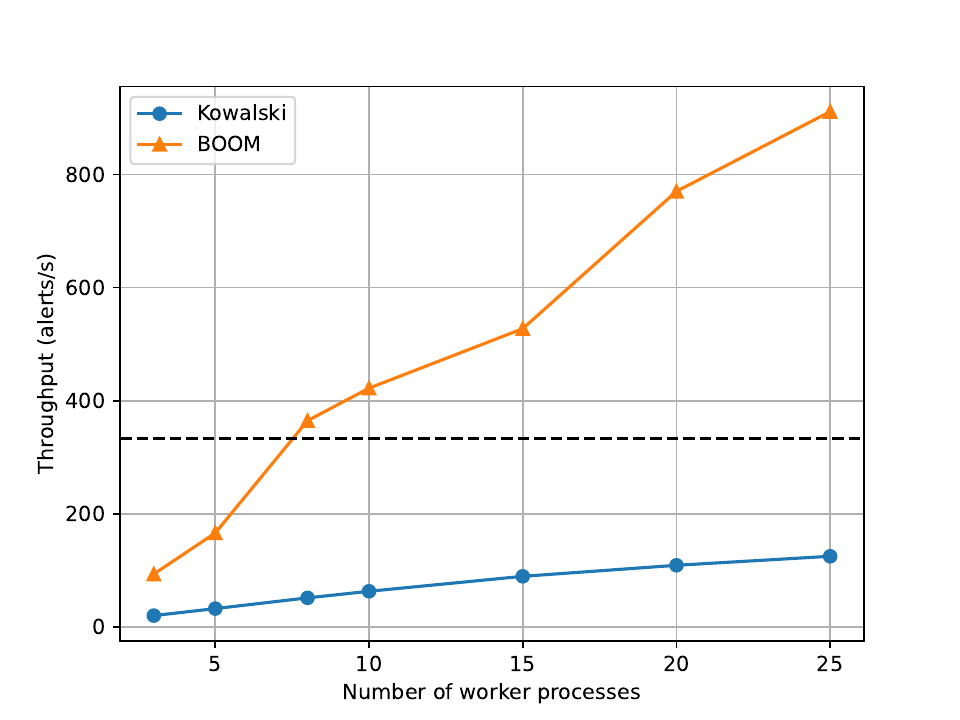}
    \caption{
        Scalability testing results for BOOM and Kowalski.
        The dashed horizontal line represents the average alert production
        rate of the Rubin observatory: 10,000 alerts for every 30-second exposure,
        or $\sim 333$ alert/s.
    }
    \label{fig:scaling}
\end{figure}

Figure~\ref{fig:scaling} shows throughput testing results for both \boom
and \kowalski in terms of alerts processed per second
versus the number of worker processes.
In addition to its increased throughput, \boom performs better as more
computing resources are added, i.e., it has better scaling characteristics.
Overall, this shows that \boom will make better use of
available computing power,
and is able to handle the Rubin alert stream with as little as 10
worker processes.
As mentioned in section \ref{sec:enrichment}, current development efforts
are now looking at the integration of new features to the enrichment worker
(parametric lightcurve fitting, new ML models, etc.), as well as improving
the support for GPU-based compute. While these introduce additional compute,
early results hint at $\sim 10 \times$ faster processing using hardware
acceleration, further improving throughput.

\texttt{BOOM}'s memory footprint is also greatly reduced,
using a maximum of \boommaxmem~GB compared to \texttt{Kowalski}'s \kowalskimaxmem~GB
for the case with \refconfignworkers \space workers,
which roughly corresponds to the point at which \boom crosses the Rubin
alert rate threshold.
Note that this value is the memory used by the workers alone
and doesn't include that used by Kafka, MongoDB, or Valkey.
% Regarding storage requirements, if we extrapolate this benchmark database's
% total size to 10 years of 10 million alerts per night as expected by LSST,
% we will require \boomstoragetenyrlsstpb~PB of storage.
% However, that is assuming all data is stored for all time, which is subject
% to revision based on user needs.

\subsection{Hardware Requirements}

\boom requires at least the following processes to run: one for each worker type, one for the main process, one to consume alerts from \texttt{Kafka}, one for \texttt{Valkey}, one for \texttt{MongoDB}, and one for its \kafka output. Since it operates in a fully asynchronous context, it can function with less physical threads than there are processes.
Memory requirements depend primarily on alert storage and job scheduling. By default, \valkey will stores up to 15,000 alerts using $\sim 1$ GB of memory, while other job scheduling items consume negligible amounts of memory. This maximum alert number can be adjusted downward to reduce \texttt{Valkey}'s memory footprint. \mongo also benefits from caching, and similarly its memory allocation can be capped at startup.

From Figure~\ref{fig:scaling}, we have shown that $\sim \refconfignworkers$ total workers are sufficient to run the benchmarked version of \boom at the LSST scale.
Therefore, we recommend running \boom on servers that have at least this number of threads to execute the software itself,
and a matching number of threads to run \mongo and \valkey efficiently.

\subsection{LSST Alerts Storage Requirements}

On Caltech's production instance we see that with LSST's current schema, 10M alerts -- which is LSST's expected alert count for a single night -- occupy 7.67 GB on disk (accounting for MongoDB's built-in compression, as otherwise the data would occupy $\sim$\,4$\times$ that volume). Therefore, if we assume LSST will observe $\sim$300 nights a year, we expect the LSST alert collection to use 2.3\,TB a year, and 23\,TB for the total length of the survey (10 years).

Estimating storage requirements for the object-level collection (\texttt{LSST\_alerts\_aux}) is not straightforward, as no representative dataset exists yet. To date, most alerts emitted by LSST originate from the deep drilling fields, which produce few distinct objects but hundreds of detections per source per night. This regime differs significantly from the wide-fast-deep survey mode under which LSST will operate for its 10-year lifespan, where most objects will receive at most 2 observations per night with gaps of $\sim$5 or more days. On the other hand, LSST's greater depth allows it to continue detecting objects at fainter magnitudes than ZTF can, extending their effective observation window. These competing effects — lower cadence but greater depth — make it difficult to directly extrapolate per-object storage requirements from ZTF data.

We can still attempt a rough order-of-magnitude estimate by scaling from existing ZTF data, though the uncertainties in this extrapolation are substantial. We assume that LSST's greater depth does not compensate for its $\sim5\times$ lower cadence — that is, individual objects accumulate \emph{at most} as many detections as their ZTF counterparts, and thus occupy no more disk space per object. This is a conservative assumption: in practice, LSST's greater depth may extend detection windows enough to partially offset the cadence difference, which would increase per-object storage beyond this estimate. Over its $\sim$8 years of operations, ZTF has produced 880\,M alerts from 222\,M unique objects, occupying 1.4 TB on disk (an alerts-to-objects ratio of 4:1). Assuming LSST emits 10\,M alerts per night over $\sim 300$ nights per year, it would generate 3\,B alerts annually, corresponding to 750\,M unique objects per year and $\sim7.5$B over 10 years. Scaling accordingly:
\[
\frac{1.4 \times 7500}{222} \approx 47.3 \text{ TB}
\]
where 7500=7.5\,B/1\,M is the ratio of expected LSST unique objects to ZTF's.

As for the disk space occupied by the alert cutout data, this all depends on what a BOOM instance can afford to store on disk. Images are by far the most storage hungry data product, and storing even just 10\,M uses $\sim 645$\,GB. Thankfully, image cutouts can easily be retrieved directly from the LSST data science platform APIs, and are never used to filter on images (instead, ML scores or features may be computed from the images as we get alerts, and these quantities are what one may query on). So, here, we plan to only keep up to N hours of alert cutouts in MongoDB, where N can be tuned to fit the disk space available on a deployment, and needs to be equal or superior to 80 hours (the planned image embargo period for new LSST images, which means one cannot retrieve cutouts from images taken in the past 80 hours and these need to be sourced from alert packets instead). The cutout API of BOOM and its public facing extension \texttt{Babamul} will be equipped to serve images sourced from the LSST data science platform APIs when these are not available in MongoDB anymore. With a retention period of 7 days (which is the value currently used by the Caltech \texttt{BOOM} instance), cutouts will never need more than $\sim 4.5$ TB.

Summing up the data requirements for the alert collection (23\,TB), object collection ($\le$47\,TB, order-of-magnitude estimate) and ``in-hot`` cutout collection (4.5\,TB with a 7-day retention period), BOOM's LSST storage requirement is roughly $\sim$75\,TB at the end of the survey's planned 10-year lifespan - likely an overestimate, given the conservative assumptions underlying the object-level figure. We emphasize that these figures are intended as planning estimates rather than precise predictions; representative wide-fast-deep survey data will be required to refine the object-level storage characterization.

\subsection{Deployment Approaches}

\texttt{BOOM}'s software repository is currently configured to automatically
deploy via GitHub Actions, running services in containers with Docker Compose.
This configuration is fully self-sufficient,
including services for \mongo, \kafka, \valkey, and of course \texttt{BOOM}'s
workers and API server, with the option of enabling the \babamul feature.

\subsection{Integration with \skyportal}

The purpose of filters implemented in brokers like \boom is to greatly limit the number of alerts that may correspond to astrophysical phenomena of interest for a given user. However, once filtered, the alerts need to flow to another system to be vetted and for actions on them to be taken. As mentioned in the introduction, this is done in TOMs, or marshals.

We have integrated the output of \boom filters within \skyportal, where user ``groups'' may have the ownership over one or many filters. While for some configurations all candidates are automatically saved as sources to a group, e.g. for automated triggering of spectroscopic follow-up, often users manually vet these filtered candidates further through a process known as candidate scanning, before proceeding with follow-up observations.

The candidate interface displays contextual image cutouts from ZTF and other relevant surveys, light curves, astrometric and photometric metadata (e.g., coordinates, cross-matches with the Transient Name Server), and direct links to external resources. Users can efficiently review this information to identify sources of genuine astrophysical interest and selectively save them for follow-up.

While \skyportal naturally allows for multiple alert streams, and so no substantial changes have been required to allow for scanning alerts that have flowed from \boom to its database, an entirely new UI framework has been developed to facilitate filter building.

\begin{figure*}[t]
    \centering
    \includegraphics[width=6.5in]{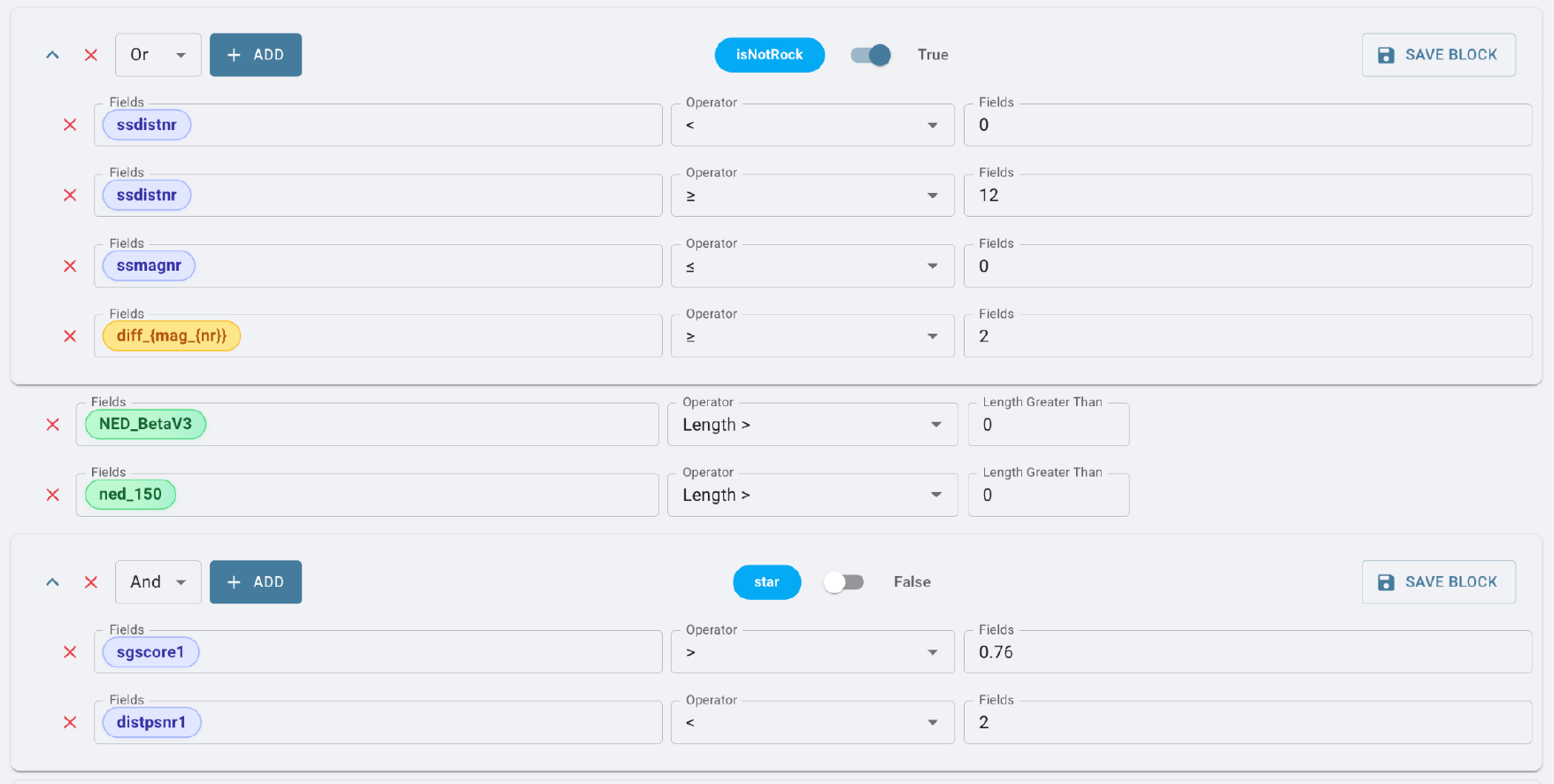}
    \caption{Filter Builder}
    \label{fig:filter_interface}
\end{figure*}

\subsection{Filter building user interface}
\label{filter_building}

To enable scientists to fully leverage the features of BOOM, we urgently need tools to facilitate the development of astronomical alert filters, by facilitating knowledge sharing and reusability, while greatly simplifying the design of such filters. To address this need, we have developed a visual block-based system that enables scientists to build filters through an intuitive form-based interface. Each filter can be exported as an independent module and later re-imported as a building block within more complex pipelines. By fully abstracting the underlying database-specific query language required to run such pipelines, we hope to redirect scientists’ efforts and attention to the higher-level decision-making and design required to successfully execute their science program.

As illustrated in Figure~\ref{fig:filter_interface}, the interface supports both basic and advanced use cases. Filters are constructed as combinations of conditions under a logical operator (AND/OR), which can be saved as reusable blocks. For instance, a block may evaluate whether a source is a star, and such blocks can then be incorporated into larger, more complex blocks. Conditions can also be applied directly to arrays or subsets of data, and an integrated LaTeX-compatible equation editor enables seamless inclusion of mathematical expressions in filters that can later be exported to be included in publications.

In addition, conditions on arrays or subsets of data are processed through a dedicated interface. After selecting an array and an operator, the user assigns a name to this array condition. Depending on the operator, the interface either presents a list of subfields from the array that has been selected, or provides a block component for constructing conditions on specific subfields. These options produce different output formats depending on the selected operator. Once saved, those custom conditions are stored in the database and are available to all users.
This modular design accelerates the development of new filters, promotes collaborative workflows, and ensures consistency between research teams. By simplifying filter creation, supporting reusability, and abstracting technical complexity, the system enhances both the efficiency and scientific rigor of astronomical alert processing through our broker.

\subsection{In preparation for LSST: Joint ZTF + DECam program}
\label{science_validation}

The DESI Transients Survey (DTS; Prop ID: 2025A-729671; PI Palmese) includes a DECam wide field survey that observes ${\sim}100$ square degrees of sky in current Dark Energy Spectroscopic Instrument (DESI) tiles as part of the larger DECam DESI Transient Survey (2DTS) \citep{palmese_desirt_2022, hall_decam_2025, hall_desi_2026}. DTS observes in the $gri$ bands down to a depth of $r\approx23.5$ on a $3$ day cadence with the goal of producing high quality light curves for thousands of extragalactic transients at a $z > 0.2$ or a peak brightness of $r\approx20.5$. As part of the 2DTS experiment, ZTF has also begun observations with a daily cadence of the same fields in the sky, offering intra-night observations at an even greater cadence.

\begin{figure*}[t]
    \centering
    \includegraphics[width=6.5in]{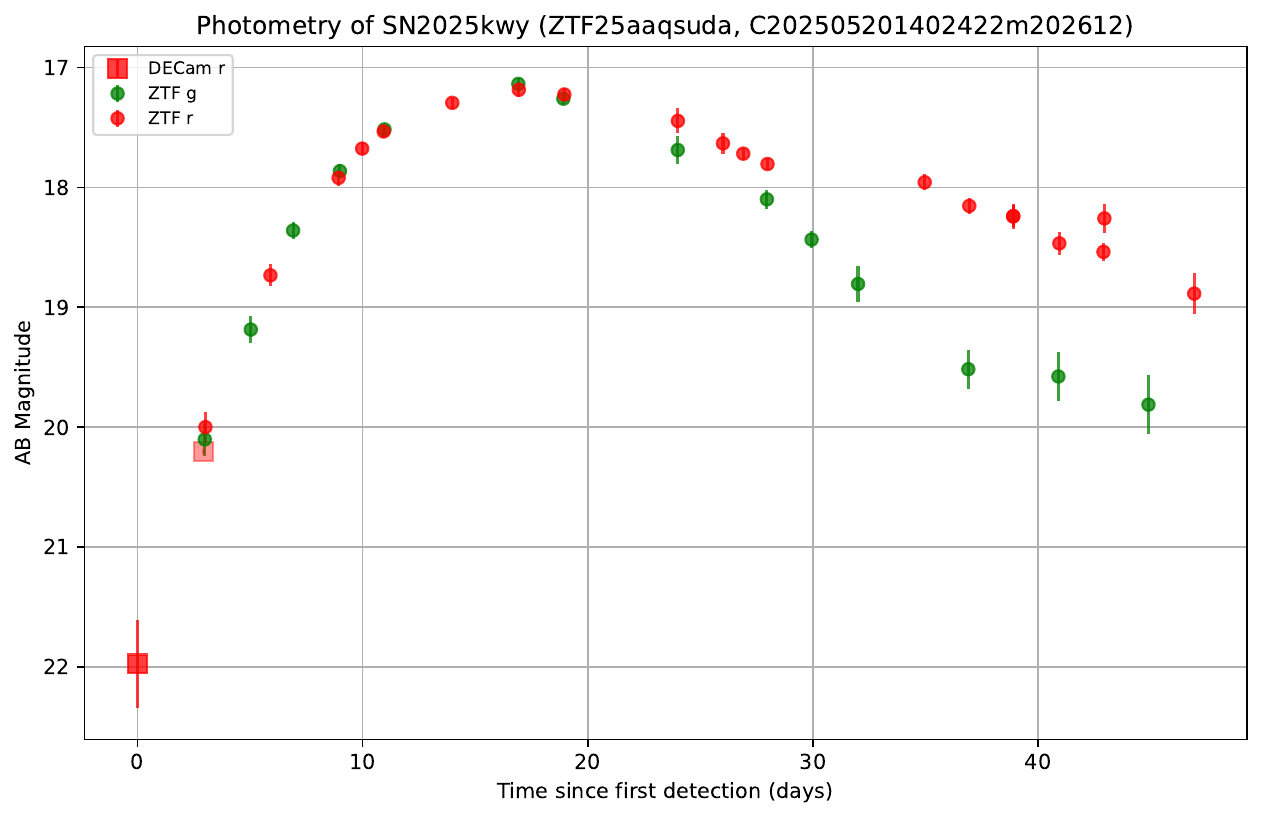}
    \caption{
        Alert photometry of SN 2025kwy, a young supernova candidate first detected by DECam and later observed by ZTF (C202505201402422m202612 / ZTF25aaqsuda).
    }
    \label{fig:ztfdecamsn2025kwy}
\end{figure*}

DTS uses the Saccadic Fast Fourier Transform (SFFT) algorithm developed in \citet{hu_image_2022} to enable fast and accurate difference imaging to identify transient alerts, see \citet{Cabrera_2024,Hu2025} for further details on the full analysis pipeline. The transient alerts are then processed with a real-bogus convolution neural network to separate unlikely artifacts such as cosmic rays. The pipeline then performs a cross-match of the alerts with Gaia DR3 to remove known stars \citep{vallenari_gaia_2023}. Finally, a match to the Legacy Survey star-galaxy catalog \citep{liu_morphological_2025} is performed to remove any remaining stellar alerts based on archival source's morphology in Legacy Survey imaging \citep{dey_overview_2019}. The remaining transients are then packaged into Alerts and sent out in a Kafka stream. The hand-selected transients, based on a visual lightcurve inspection, are then reported to TNS. This program offers a unique prelude to the issues of matching alert streams between a relatively smaller telescope such as ZTF and larger telescope like LSST. The observational depth of DTS is $\sim$\,3\,mag deeper than ZTF, comparable to the $\sim$\,4\,mag difference LSST will have.

Over the course of a 3-day experiment conducted in May 2025, ZTF and DECam observed spatially coincident fields (Prop ID: 2025A-898110; PI Ahumada). To test \texttt{BOOM}'s abilities to handle alert streams with vastly different depths, a Kafka stream with ZTF-styled \texttt{Avro} alerts was developed for DTS, and the resulting alerts were emitted and processed alongside ZTF's by \texttt{BOOM}. We observed 207 ZTF objects with matching DECam objects; this relatively small number is due to the pre-filtering used by DECam before sending alerts, as mentioned above. In \texttt{BOOM}, these were matched by the \texttt{Alert} worker, and a simple user-defined filter looking for ZTF alerts with matching DECam transients was implemented, which identified these 207 multi-survey candidates and sent these to \texttt{BOOM}'s \kafka cluster, and thereafter read and ingested by a dedicated \skyportal instance.

Although none of the transients discovered in the course of the short experiment were found solely due to the ZTF and DECam joint filtering synergy, we were able to validate the cross-survey matching and cross-survey filtering capabilities of the software with real data. Amongst the transients observed by both surveys, we highlight one particular object, SN 2025kwy. Later classified as a SNIa, it was first detected by DECam at 21.98\,mag in $r$-band, followed by another $r$-band detection 3 days later at 20.2\,mag, followed the same night by ZTF detections in $r$-band and $g$-band; it was observed $\sim$\,daily by ZTF thereafter. While DECam data alone was sufficient to constrain the transient's rate of evolution at early times---using multiple detections in the same band---within 3 days of the first detection, this would not be possible with LSST's current plan to take observations in different filters within a night and then return to a field within a week. Therefore, to simulate what may be expected from the LSST alert stream, we sub-sampled from the DECAM lightcurve and only kept the first---and fainter---detection, 3 days before ZTF's first observation. This leaves us with a ZTF + LSST joint-stream example as illustrated in Figure~\ref{fig:ztfdecamsn2025kwy}, highlighting the synergy between the 2 surveys: fainter pre-detections by LSST, followed by a higher cadence, well-sampled lightcurve from ZTF; naturally in the LSST case, the light curve would also be filled in with further photometry from other passbands. In this scenario, the multi-survey support lets us put clear constraints on the rate of evolution of the transient at early times ($\sim 0.65$ mag/day for the first 3 days), resulting in its selection by fast-transient or young-supernova user-defined filters. Moreover, \boom had cross-matched the transient with the \texttt{NED LVS} catalog, adding additional information about its distance, host, and absolute magnitude, all which can be used as additional constraints in user-defined filters.

\subsection{Live LSST alerts and public interface}
\label{production}

As of February 25th 2026, LSST alerts are now publicly available through all Rubin alert brokers, including BOOM's public facing interface: \texttt{Babamul}\footnote{\url{https://babamul.caltech.edu}}. \texttt{Babamul} serves as a general-purpose alert broker for the U.S. and international astronomy communities taking advantage of the multiple surveys currently online, by providing a number of public streams supporting a variety of science cases through \kafka topics, based on the many features and flags added by \texttt{BOOM}'s workers. In fact, \babamul can be enabled in any \boom instance with a simple feature flag. When enabled, an additional processing step occurs for every batch of alerts processed in the enrichment worker. There, we:

\begin{itemize}
    \item First, filter out low-quality alerts and solar-system candidates. The filter conditions are survey-specific and will evolve over time especially for LSST. Moreover, we hope to bring back SSO candidates into their own dedicated Kafka streams, at a later date.
    \item Secondly, we categorize each alert to associate it to a single kafka stream. \babamul aims to deliver disjoint (guaranteeing no duplication) streams that redistribute the entire alert stream to the community, split into its different parts. For example, a ZTF alert with an LSST match and matching to a known stellar object will be assigned to the \texttt{babamul.ztf.lsst-match.stellar} Kafka stream.
    \item Alerts are bundled in an Avro package that contains: all metadata about the current alert, the features computed by the Enrichment worker, a lightweight lightcurve schema, matches with other surveys and their lightcurves, but no image cutouts. This lets us trim down on the alert stream size significantly, necessary to support many users in parallel. The information bundled in \babamul alerts are more than sufficient for downstream users to run a filter layer of filtering to narrow down the number of relevant candidates drastically. For those that are left, the API may be used to retrieve additional data products (crossmatches with other catalogs, all lightcurve metadata, and images).
    \item Last but not least, the Avro-serialized alerts are sent to their associated Kafka stream, configured to enable compression at the Kafka-level (which offers higher compression rates and throughput---for both server and client---than when enabled at the Avro-level).
\end{itemize}

\babamul ships with a dedicated Python client\footnote{\url{https://pypi.org/project/babamul/}} to facilitate accessing the alert streams. In this client, Pydantic is used to deserialize Avro packets into Python classes, enabling type-safety checks, auto-completion in IDEs, and easier integration with Agentic coding tools. Moreover, by treating each alert as a full-blown Python class, this lets us add various helper functions and methods to facilitate using the data products or fetching additional information from the API. Example use-cases for the client are provided as Jupyter notebooks\footnote{\url{https://github.com/boom-astro/babamul/tree/main/examples}}.

\texttt{Babamul}'s web application serves as the entry point to its Kafka cluster, Restful API, Python client, and help sections. But, its main purpose is to provide visualization tools for each objects, as well as tools to search for them in an intuitive manner. So far, individual object pages and ID-based searches have been implemented, while coordinate and property based searches---all already available through the API---are still in the works. Also, a dedicated SkyPortal extension is actively being developed. Based on the Python client, it will let members of the community filter and ingest revelant alerts to their own TOM and marshal instances.

Since LSST alerts have been made public, many transients of interest were identified using \boom or \texttt{babamul} (\cite{2026TNSAN..43....1H}, \cite{2026TNSAN..44....1K}, \cite{2026TNSAN..68....1J}), and reported to the community. Also, ZTF's own SkyPortal instance---Fritz---is now equipped with the filter building UI described in section \ref{filter_building}, and ZTF collaborators have designed filters that run on the stream in real-time, with Fritz then ingesting the identified candidates to enable various candidate scanning workflows, as done previously with Kowalski (but with the addition of multi-survey information).

\section{Conclusions}
\label{sec:conclusion}

In conclusion, the data processing system we have described
above---\texttt{BOOM}---is now operational on ZTF and well-positioned to scale to LSST.
Designed to deliver real-time filtering of incoming alerts, it also offers a queryable archive of all ZTF and LSST alerts throughout the surveys' operational lifetime. This archive will enable retrospective analyses and is structured to support future batch-processing capabilities, facilitating large-scale, post-facto scientific investigations. More broadly, \boom empowers researchers with a flexible, scalable platform that naturally allows for brokering multiple surveys simultaneously which enables for the extraction of the maximum scientific value from current and future survey.

\texttt{BOOM}'s ability to empower science users to select interesting transient candidates in both LSST and ZTF alert streams has already enabled multiple Transient Name Server AT reports since Feb 25th 2026, and we are eager to see the many scientific publications that will use it -- or its public interface \texttt{Babamul} -- as their discovery engine.

In the future, we have a number of critical developments we plan for the platform, mostly focused on facilitating alert filtering. Namely, we propose the development of software dedicated to re-running these complex filters after the fact, to validate their capabilities, purity, and to estimate the rates at which we can expect automated ToOs for the targeted follow-up instruments.
Re-running filters is already possible through \texttt{BOOM}'s HTTP API and through the filter-building UI, but enforcing validation of the filter results before proceeding to submission for real-time operations---which requires additional development work---is lacking. This way, any iteration of a given filter would automatically come with associated statistics, building a strong baseline and point of reference as we iterate to improve any survey's results. Moreover, we are looking forward to the integration of additional machine learning models and lightcurve fitting tools, all supported by hardware acceleration to improve throughput even further, and facilitate new discoveries.

\section*{acknowledgments}
\label{sec:acknowledgments}

We acknowledge support from the National Science Foundation grant number 2432476, ``Delivering Open, Accessible and Collaborative Infrastructure Enabling Multi-Messenger Astrophysics".

Based on observations obtained with the Samuel Oschin Telescope 48-inch and the 60-inch Telescope at the Palomar Observatory as part of the Zwicky Transient Facility project. ZTF is supported by the National Science Foundation under Grants No. AST-1440341, AST-2034437, and currently Award AST-2407588. ZTF receives additional funding from the ZTF partnership. Current members include Caltech, USA; Caltech/IPAC, USA; University of Maryland, USA; University of California, Berkeley, USA; University of Wisconsin at Milwaukee, USA; Cornell University, USA; Drexel University, USA; University of North Carolina at Chapel Hill, USA; Institute of Science and Technology, Austria; National Central University, Taiwan, and OKC, University of Stockholm, Sweden. Operations are conducted by Caltech's Optical Observatory (COO), Caltech/IPAC, and the University of Washington at Seattle, USA.

M.W.C. acknowledges support from the National Science Foundation with grant numbers PHY-2117997, PHY-2308862 and PHY-2409481.



\bibliographystyle{aasjournal}
\bibliography{references} % if your BibTeX file is called example.bib

\label{lastpage}
\end{document}